\documentclass[12pt]{iopart}
\usepackage{graphicx,color,amssymb,amsfonts,hyperref,soul}
\usepackage[usenames,dvipsnames]{xcolor}
\usepackage[frenchb,english]{babel}
\definecolor{nblue}{rgb}{0.3,0.3,1.0}
\definecolor{ngreen}{rgb}{0.2,0.7,0.2}
\definecolor{nred}{rgb}{0.9,0.1,0}
\definecolor{npurple}{rgb}{0.8,0.2,0.8}
\definecolor{golden}{rgb}{0.8,0.6,0.1}
\definecolor{nsilver}{rgb}{0.3,0.4,0.5}
\definecolor{nbrown}{rgb}{0.8,0.4,0.15}
\definecolor{nrose}{rgb}{0.7,0,0.35}
\definecolor{nviol}{rgb}{0.5,0,1.0}
\definecolor{nazur}{rgb}{0,0.35,0.7}
\definecolor{nchart}{rgb}{0.2,0.4,0}

\newcommand{\vio}{\color{nviol}}
\newcommand{\blk}{\color{black}}
\newcommand{\blu}{\color{nblue}}

\newcommand{\gold}{\color{golden}}

\newcommand{\beq}{\begin{equation}}
\newcommand{\eeq}{\end{equation}}
\newcommand{\bqa}{\begin{eqnarray}}
\newcommand{\eqa}{\end{eqnarray}}

\begin{document}

\title[]{Reply to Gillis's ``On the Analysis of Bell's 1964 Paper by Wiseman, Cavalcanti, and Rieffel'' }
\author{Howard M. Wiseman}
\address{Centre for Quantum Dynamics, Griffith University, Brisbane, Queensland 4111, Australia}
\author{Eleanor G. Rieffel}
\address{QuAIL, NASA Ames Research Center, Moffett Field, CA 94035}
\author{Eric G. Cavalcanti}
\address{Centre for Quantum Dynamics, Griffith University, Gold Coast, Queensland 4222, Australia}
\address{}

\begin{abstract}
We address Gillis' recent criticism of a series of papers (by different 
combinations of the present authors) on formulations of Bell's theorem. 
Those papers intended to address an unfortunate gap of communication 
between two broad camps in the quantum foundations community that we 
identify as ``operationalists'' and ``realists''. Here, we once again 
urge the readers to approach the question from an unbiased standpoint, 
and explain that Gillis' criticism draws too heavily on the philosophical 
inclinations of one side of that debate -- the realist camp. As part of 
that explanation we discuss intuition versus proof, look again at 
Bell's formalizations of locality, and correct misstatements by Gillis 
of our views, and those of Bell and Einstein. 
\end{abstract}

\section{Introduction}

Gillis's ``On the Analysis of Bell's 1964 Paper by Wiseman, Cavalcanti, 
and Rieffel'' \cite{Gil15} refers to three papers 
\cite{Wis14b,WisCav16,WisRie15} that discuss opposing views of 
Bell's theorem and attempt at least a partial 
reconciliation.\footnote{Wiseman, the only author on all three of 
these papers, was asked to report on Ref.~\cite{Gil15}, but declined 
to take any part of the refereeing process because of a 
clear conflict of interest.} 
Wiseman's original paper \cite{Wis14b} carefully examined assumptions
behind the two predominant views, which he termed the 
operationalist and realist views, and performed a detailed textual analysis
of papers of Bell's, particularly his 1964 paper \cite{Bell64} to determine 
what was proved, and under what assumptions, at that time. 
His conclusion \cite{Wis14b, Wis14a}
was that Bell proved two different theorems, one in 1964 and one in 1976, 
the first corresponding the view of Bell's theorem taken by the 
operationalist camp, the second to the view of the realist camp.

Essentially everyone agrees that the two theorems are mathematically correct.
But there is huge disagreement as to whether the first is (a) 
of any interest, and (b) due to Bell. At the heart of the disagreement is 
the meaning of various localistic phrases in the writings of Bell's and others.
In this response, we will discuss only analyses relevant
to Gillis's arguments. For a more general discussion, please see
\cite{Wis14b} and the two subsequent papers \cite{WisCav16,WisRie15}.

Wiseman's paper has achieved some of its aims toward better communication
between the two camps. For example, many operationalists are now aware that
there is a different and interesting Bell's theorem from the one they knew.
There remains staunch opposition in some corners of the realist
camp to recognizing any legitimacy to the operationalist view
\cite{Mau14,Nor15,Gil15}, though some of the most vocal critics have 
come part way along the path Wiseman laid out \cite{Nor15}.
The issue that raises the most passion, and over which there 
is least agreement, is the content of Bell's 1964 paper. 

Since this question is a purely
historical matter, one might ask how much effort should be spent
by the physics community in resolving it. We believe there are strong
reasons to resolve this issue. One of the key claims by some of the
realist critics, including Gillis, is that the correct notion of locality 
has been clear enough since the 1935 paper of Einstein, Podolsky and Rosen
\cite{EinPodRos35}, with the implication that other formulations of Bell's theorem are not worth considering.  We
believe that this view is not only historically incorrect, but also holds
researchers back from understanding and contributing to ongoing research  
such as \cite{CavLal14,WooSpe12,WisCav16}
that are elucidating the relation of various notions of causation 
to quantum mechanics. For this reason, addressing these claims
has impact beyond the `merely' historical.

Here, we respond specifically to arguments contained in 
Gillis's ``On the Analysis of Bell's 1964 Paper by Wiseman, Cavalcanti, 
and Rieffel'' \cite{Gil15}.  Because his paper is less an analysis of 
our papers than a strongly stated defense of the realist stance, we will 
at times, of necessity if 
briefly, review material that is already contained in our previous papers.
In none of our papers \cite{Wis14b,WisRie15,WisCav16} do we argue 
for either the realist 
or the operationalist camp -- all three works recognize the value of both
Bell's Theorems and aim to
explain why different versions of Bell's theorem are favoured by each camp 
and to bridge communication gaps between the camps. 
Because Gillis's paper comes from the realist point of view, it may appear
that in this paper we are taking the operationalist side. That is not
the case. Rather, we are trying to explain what operationalists will find
unconvincing about his arguments.
Gillis engages surprisingly little with the arguments in our
papers, focusing on only a few sections of \cite{Wis14b}, discussing only
one passage in \cite{WisRie15}, and making no specific mention 
of arguments in \cite{WisCav16} or its new formulation in terms of 
causation that provides a means to bridge the gap between the two camps. 
Thus, there is less to respond to in terms of specific disagreements 
with our arguments than might be expected. 

In response to Gillis's paper, we discuss intuition versus proof, 
look again at Bell's formalizations of locality, and correct misstatements 
by Gillis of our views, and those of Bell and Einstein. 
Before getting to those deeper issues, we address a question of terminology.

\section{Objection to WCR terminology}

We object to the use of WCR for what Wiseman called 
the operationalist viewpoint on the content of the 1964 paper 
-- it incorrectly gives us credit for a widely held view, one
that Bell himself complained \cite{Bel81} was the reading 
``\blu almost universally reported\blk."\footnote{We will use
\blu blue \blk for quotes from Bell, 
\gold gold \blk for quotes from Gilles, 
\vio violet \blk for quotes from our papers.} 
To name it after us is highly misleading, and is not 
the ``\gold neutral terminology\blk" that Gillis claims it is. 
By making it appear to be a disagreement with us, rather
than with the views of a large community, he does not have to
wrestle with why a large number of people hold that view and provide
an alternative explanation to Wiseman's as to why they do.
We will continue to refer to the operationalist view, 
but if Gillis and others would like ``\gold more neutral\blk'' terminology, 
we recommend they use AUR, for ``\blu almost universally reported\blk," 
rather than WCR. 

Gillis also uses WCR to refer to a personified
 entity that will ``\gold formalize\blk," ``\gold insist\blk," and 
``\gold fail to distinguish\blk." 
Such an entity is
entirely hypothetical. Prior to this paper, the three of us had not 
published together, so any such statement would need at least two
citations. While in some cases his use of WCR may reflect views all three
of us share, in many cases he is referring to material in Wiseman's solo
paper, and sometimes to material in a two-author paper, but often 
his hypothetical entity, as we will illustrate, espouses views that 
none of us hold.

\section{Intuition versus mathematical formulations}
\label{sec:viewsOnformalism}

One of the reasons we expect that readers who do not already hold
Gillis's views will not find his arguments convincing is that
he makes no distinction between Bell's intuitive statements and his
more formal statements  and proofs.
Wiseman's paper \cite{Wis14b} was concerned with what Bell did, and did 
not, prove in the 1964 paper, not his intuitions, however interesting, 
at the time. Had Gillis made a clear distinction between 
Bell's intuitive views in 1964 and what he had succeeded
in formalizing and proving by that time, we expect 
many of his disagreement with \cite{Wis14b} and 
the subsequent papers \cite{WisRie15,WisCav16} would vanish.

Gillis quotes from Bell's `prequel' paper (published in 1966~\cite{Bel66}) 
``\blu... there are features which can reasonably be desired in a hidden 
variable scheme. The hidden variables should surely have some spacial 
significance and should evolve in time according to prescribed laws. These 
are prejudices, but it is just this possibility of interpolating some 
(preferably causal) space-time picture, between preparation of and 
measurements on states, that makes the quest for hidden variables 
interesting to the unsophisticated\blk" as support for his claim that
``\gold prior to the 1964 paper Bell had already described a concept 
essentially equivalent to what he called 'local causality' in 1990.\blk" 
In doing so, he confuses early intuitions with later formalizations. Bell
himself was clear on the distinction, seeing both the desirability and the
difficulty of formalizing intuitive notions and being rightly cautious about
the potential for incorrect formalizations. 

Wiseman has never disputed the plausibility of Bell having always 
had a heuristic localistic notion which he understood OQM 
(Orthodox Quantum Mechanics) to lack, but he only formalized 
that notion, as local causality, in 1976~\cite{Bel76}. What
he offered in 1964 as a definition  was a statement of locality that requires 
a little interpretation to be formalized, but is quite different from local causality. 
By one reading, locality as Bell used it in 1964 could only be 
applied to deterministic theories, in which case it cannot even
be applied to quantum theory, so cannot be used to make
any claim about the locality or nonlocality of OQM.
More charitably (which has always been Wiseman's approach) it could 
be applied to probabilistic theories but (unfortunately for Bell) 
it still does not imply that OQM is nonlocal. 

As a side note, Gillis's hopes the vague intuitions 
Bell gives in the `prequel' paper provide a notion of locality
from which predetermination could be derived from predictability. 
But the quotes all 
discuss the potential localistic properties (which Bell thought would be 
desirable) of {\em hidden variable theories}. So this notion, even if had 
been carefully formulated, which it had not, would clearly not have enabled 
one to {\em derive} the existence of hidden variables from 
any quantum correlations. If this is what Bell meant by locality, it would 
make his suggestion that OQM is nonlocal not merely false, but foolish 
(to steal a phrase used by Bell~\cite{Mau14}). 

In suggesting that the concept of local causality was there all the time, 
Gillis fails to recognize the formidable intellectual effort, 
starting with Bell's 1971 paper~\cite{Bel71}, then the Clauser-Horne 
paper of 1974~\cite{CH74}, then Bell's 1976 paper~\cite{Bel76} and the 
subsequent conversation about free will~\cite{ShiHorCla76,Bel77}, 
to develop the formalization of local causality. Furthermore, the
assurance with which Gillis asserts that this concept is the way
to capture these intuitive notions runs counter to Bell's own thinking
on this subject. 
Even in 1990, after thinking about these issues for decades, 
in his final essay on the subject, La Nouvelle Cuisine \cite{Bel90b}, 
a main theme is 
``\blu the problem of formulating [cause and effect] sharply in 
contemporary physical theory.\blk" 
In this same essay, notice how tentative he is at the very point at
which he introduced local causality even at this late date: 
``\blu it is precisely in cleaning up intuitive ideas for
mathematics that one is likely to throw out the baby with the bath
water. So the next step should be viewed with the utmost suspicion: A
theory will be said to be locally causal...\blk" 
When interpreting Bell, it is critical to recognize both the value
he placed on correct formalization of intuitions and his extreme
caution when attempting to do so.

\section{What Bell meant by `locality'} 

Gillis make an interesting suggestion, on page 7, that what Wiseman has 
taken to be Bell's definition of locality, 
\begin{quote} \blu
It is the requirement of locality, or more precisely that the result 
of a measurement on one system be unaffected by operations on a 
distant system with which it has interacted in the past, that 
creates the essential difficulty. \blk
\end{quote}
could instead be read as stating what Bell took to be a {\em consequence} 
of locality (whatever that may be). We admit that is a possible reading. 
However, his phrase  ``\blu more precisely that\blk" most naturally 
indicates a definition. If he had meant merely ``\gold a particular 
consequence of locality\blk," as Gillis suggests, then Bell 
should have said ``more particularly that." Whatever the case may be,
it doesn't help move Gillis's general argument forward. Bell does not 
formalize any broader concept of locality that could play the role that 
Gillis wants it to play.  

Gillis suggests a variety of vague statements to describe what 
he claims to be Bell's views of locality in 1964 such as ``\gold no 
action-at-a-distance\blk,'' ``\gold continuous, subluminal propagation 
through space\blk,'' and ``\gold No Superluminal Effects (NSE)\blk" 
all of which he presumes mean roughly the same thing
as local causality as Bell introduced it in 1976. 
A phrase like ``\gold no action-at-a-distance\blk'' 
means next to nothing in itself. It cannot be used to derive any conclusions 
until it has been formalized, and cannot be used to reach agreement
on fundamental issues with  
those holding different intuitions about what these phrases mean. 

With respect to Bell's assumptions and proofs in the 1964 paper, 
nothing in Gillis's paper is likely to dissuade readers of
Ref.~\cite{Wis14b} from its key claims, briefly, 
\begin{enumerate}
\item Bell does not provide, in 1964 or earlier, a formalization of locality that  allows a rigorous derivation of his result without 
assuming predetermination of outcomes (or something like it). 
\item This was not simply because Bell could not be bothered to formalize 
something obvious. Rather, it was a major effort over several years 
and several authors, to get to local causality as Bell defined it in 1976. 
\item Bell shows no appreciation in 1964 of the nuances of Einstein's 
earlier work. Bell does not engage at all with the technical content
of the EPR paper~\cite{EinPodRos35}, and he also misses the crucial dual 
assumptions in Einstein's 1949 scientific 
autobiography~\cite{Ein49}; see also Ref.~\cite{WisRie15}.
\item Bell is very clear about his assumptions in 1964, stating 
four times that his theorem assumes predetermination. He had an 
{\em informal} argument in favour of considering this assumption, but 
he knew that he did not have a {\em formal} notion of locality 
that would do the job without the assumption of predetermination.
\end{enumerate}

On pp.~8--9, Gillis attempts to explain why Bell phrased his paper in terms of settings
rather than a more general notion of locality. He suggests that ``the
reason is that it is the one aspect of the experimental arrangements envisioned that
can be placed unambiguously outside the past light cone of the second measurement'', and recalls that ``Bell
was hugely influenced by Bohm's theory'', in which it is the setting of one measurement that influences the distant outcome. 

However, these arguments provide no support for the idea that Bell had a more general intuitive notion of locality; on the contrary, they can explain why Bell may have initially thought about locality primarily in terms of the settings, and why he was only able to formulate his notion of local causality at a 
later date. 

\section{Correcting misstatements of our views, and those of Bell and Einstein}

Gillis presents some material in a way
that would suggest to many readers that the material is new to the discussion,
when in fact it was already discussed in Wiseman's paper, often in greater
depth. Examples include the factorizability condition, Jarrett and Shimony's
notion of Jarrett completeness or outcome independence, and the relation
between these concepts. 

In this section, we attempt to correct Gillis's serious misstatements of 
our views, and those of Bell and Einstein.

\subsection{Quoting Bell out of context}

Gillis claims that Bell's 1964 paper~\cite{Bel64} proved 
that \gold `NSE' \blk (\gold ``No Superluminal Effects''\blk) is 
contradicted by \gold `QSP' \blk (\gold ``Quantum Statistical 
Predictions''\blk). Bell uses no term similar to NSE --- he uses 
neither the term `cause' nor the term `effect' ---  
which should raise warning bells about Gillis's analysis. Nevertheless, 
Gillis quotes Bell, towards the end of his (Gillis's) Sec.~2, 
to attempt to support his (Gillis's) interpretation of Bell:
\begin{quote}
\blu Moreover, the signal involved must propagate instantaneously, so 
that such a theory could not be Lorentz invariant. \blk
\end{quote}
However, in context, this quote actually supports the operationalist or
AUR reading: that Bell made two assumptions 
in 1964, locality and predetermination. The context is, in fact, 
provided by Gillis at the beginning 
of his Sec.~1: The immediately preceding sentence is 
\begin{quote} \blu
In a theory in which parameters are added to quantum mechanics to determine the results of individual measurements, without changing the statistical predictions, there must be a mechanism whereby the setting of one measurement device can influence the reading of another instrument, however remote. \blk
\end{quote}
Thus, when Bell  says ``\blu such a theory\blk'' he means a theory 
with predetermined measurement outcomes, not arbitrary theories. 

In the next sentence, which Gillis also quotes, Bell raises the 
possibility that (to use Gillis's term) QSP might be false. 
Bell naturally did not raise the `possibility' that determinism
might be false because all of his readers would have known that. 
In 1964 the challenge was to convince readers
that there was {\em any} point considering deterministic theories at all. 
This is why he begins in the first sentence with the ``\blu paradox 
of EPR\blk", to motivate the idea of hidden variables. Note that he does not, 
in the abstract, say that the EPR argument implies 
determinism/causality from locality. He just says 
it was ``\blu advanced as an argument [for hidden variables].\blk " 
He does not even commit to it being a correct argument, and in this 
he certainly reflects the prevailing mood at the time 
(see also Sec.~\ref{EPRBA} below).

As a second, related example, Gillis says that ``\gold Wiseman's 
interpretation of `locality' as PI is inconsistent with Bell's 
statement at the beginning of the  [1964] paper that additional (i.e.., hidden) 
variables were needed to restore locality to quantum theory.\blk"
Gillis is cherry picking here, and not quoting Bell properly. What Bell
actually says is ``\blu These additional variables were to restore to 
the theory causality and locality [2].\blk" By leaving out 
``causality" and introducing the word ``needed'',
Gillis changes the meaning in an important way. 
Revealingly, in his concluding paragraph, Gillis quotes the final
part of Bell's sentence, but puts ellipses in place of ``causality
and." In order to have any chance of convincing operationalists of
his reading, Gillis would have to explicitly deal with the appearance of 
causality in these sentences. 

\subsection{EPR and Bohm and Aharonov} \label{EPRBA}

Gillis claims that the EPR argument was well understood as an 
``\gold obvious clash between the principle of no superluminal 
action-at-a-distance (which was regarded as essential to relativity), 
and the perfect correlations between spacelike-separated measurement 
results that quantum theory predicted, but could not explain.\blk'' 
This is not, however, how EPR presented the argument. They never 
mentioned relativity and they certainly do not formulate a 
``\gold principle of no superluminal action-at-a-distance.\blk'' 
This is crucial to understand: 
{\em No physicist had precisely formalized any such principle}, 
applicable to probabilistic theories such as quantum mechanics, that Bell could 
plausibly have used in 1964 as the single assumption to derive his inequality 
(see Sec.~3.3 of Ref.~\cite{Wis14b}). We grant, as in \cite{WisRie15}, that
``one could undertake to formalise Einstein's assumptions so as to make the argument 
from predictability to determinism", but this had not been done in 1964.
Without reasonably precise definitions, it is impossible to prove anything, and Bell rightly believed that 
he had proved something~\cite{Wis14b}. 

Gillis quotes the 1957 paper by Bohm and Aharonov~\cite{BohAha57}, 
which Bell was certainly aware of prior to 1964, as evidence that 
physicists, Bell included, well understood the EPR argument. 
But the Bohm and Aharonov passage is far from a clear 
explanation of the EPR paradox. It is even less formal than Bell's presentation. 
It is merely an intuitive argument, containing no definition of locality from 
which one could draw the conclusion that OQM is nonlocal. This is to 
be contrasted with the EPR paper itself, which is very formal, with a 
complex web of necessary and sufficient conditions, albeit still not 
completely unambiguous (see Ref.~\cite{Wis13}, and the appendix in 
Ref.~\cite{Wis14b}). 

Gillis says ``\gold Given this very widespread understanding, it was 
entirely reasonable for Bell to proceed based on a brief, informal 
recapitulation of EPR.\blk'' But Gillis presents no evidence that there was 
widespread understanding of any precise formulation of an EPR-style argument 
from perfect correlations to predetermined outcomes. As Wiseman and Rieffel
note~\cite{WisRie15}, the famous 1969 paper by 
Clauser {\it et al.}~\cite{CHSH69} is non-committal about the correctness 
of the EPR argument (our p.~97), and Bell in 1971~\cite{Bel71} says 
only that perfect correlation ``\blu strongly suggests\blk'' 
predetermination (our p.~88). 
This, and much other evidence we present, supports our contention that 
Bell did not have a formal argument in mind in this part of the paper. 
Nevertheless it was indeed ``\gold entirely reasonable\blk'' for him 
to give an informal argument because this part of the paper appears 
merely as a motivation for his assumption of predetermination, and Bell 
explicitly indicates that this is the role of the EPR argument in 
his 1971 paper~\cite{Bel71}. 

It would be unfair to expect Bell's first paper to be flawless. He 
had simply not thought hard enough, or written slowly enough, in 1964, 
to be completely consistent. He had proven a brilliant new theorem and 
wanted to communicate that quickly. 
Gillis says the first paragraph of Bell's ``Formulation" is 
`\gold strained\blk' under Wiseman's interpretation. The strain is 
unavoidable because of the flaw in Bell's exposition itself, as others who
disagree with Wiseman's more general arguments have admitted~\cite{Nor15}. 
The argument Bell gives is, at best, heuristic. 
At this point in time 
Bell had himself not sorted out the difference between his intuitive
localistic notions and the notion of locality he actually uses in his theorem. 
But his {\em theorem} is not at all flawed 
because he is always clear in his 1964 paper, when stating his theorem, 
that he assumed predetermination. 

\subsection{Bell's logic when referencing Einstein} 

Rieffel and Wiseman~\cite{WisRie15} described Bell's citations of Einstein 
in his 1964 paper as interruptions of his sentences. Gillis seems
to have entirely misunderstood our argument when he says
``\gold they cannot rewrite Bell's paper in order to eliminate portions 
that conflict with their interpretation of it.\blk" 
We did not say ``\vio In this, and every other, instance, the 
reference `[2]' could be omitted from the sentence, and it would 
actually improve the grammatical and scientific clarity of the sentence,\blk"
because of any ``\gold conflict\blk'' between Bell's text and our interpretation 
of it. Rather,  we were 
making the point that one cannot, as some have attempted~\cite{Nor15}, 
read Bell's paper as if the text of the 
footnote were present in the text anywhere the ``[2]'' appears. We were thus 
inviting the reader to consider why Bell used a footnote here rather
than including this material in the main body of his text. 
To us, the answer is clear: the quotation of Einstein expresses 
a similar notion of locality to the more precise one which Bell 
introduces, and Bell was appealing to his authority to motivate it. 
If these ideas and phrases of Einstein's were absolutely crucial to 
the understanding of his (Bell's) text, it would have been an exceedingly 
odd choice on Bell's part to place them in a mere footnote. 

For the benefit of the reader, the Einstein quote in the footnote 
in question reads:
\begin{quote} 
But on one supposition we should, in my opinion, absolutely hold fast:
the real factual situation of the system $S_2$ is independent of what is done
with the system $S_1$, which is spatially separated from the former.
\end{quote}
As discussed in~\cite{Wis14b}, and in greater detail in Ref.~\cite{WisRie15}, 
Einstein himself states that an extra assumption is needed to 
conclude that OQM is incomplete 
-- he explicitly assumes that the conditional wavefunctions of the 
distant systems are part of such ``real factual situations''. Thus, even if 
Bell's 1964 notion of locality were to be equated with the Einstein 
quote, it would not necessarily be a property that quantum mechanics lacked, 
and it could certainly not be equated with local causality. 

\section{Conclusion} 
\label{sec:conclu}

Wiseman's main aim in writing \cite{Wis14b,WisCav16,WisRie15} was to 
facilitate communication 
between two interpretational camps, whom he called `operationalists' 
and `realists', by explaining carefully how and why they use 
the phrase ``Bell's theorem'' to mean different things:
his 1964 theorem (assuming locality and determinism) and his 1976 
theorem (assuming local causality), respectively.  
As Wiseman has emphasized many times, his papers address Bell's 
theorem(s), not Bell's intuitions. It is perfectly legitimate to try 
to understand Bell's intuitions, of course, but these intuitions should 
not be confused with his stated assumptions, and what he proved from 
those assumptions. 

It remains the hope of all three of us that
through careful consideration of what Bell proved, and what notions
were formalized, Gillis and others in the staunchly realist camp can 
come to a better understanding of, and sympathy for, 
the views of those in other camps. With that should come a yet deeper respect
for Bell's substantial intellectual effort over many years to
formalize localistic intuitions
as local causality. Finally, we hope that Gillis and others may also 
come to appreciate, and even
contribute to, recent efforts \cite{CavLal14,WooSpe15,WisCav16, Cavalcanti2016} to make
further advances in solving
``\blu the problem of formulating [cause and effect] sharply in 
contemporary physical theory\blk." 

\vspace{5ex}

\section*{Acknowledgements} 
EGC acknowledges support from an Australian Research Council grant DE120100559, and a grant (FQXi-RFP-1504) from the Foundational Questions Institute Fund (fqxi.org) at the Silicon Valley Community Foundation. HMW acknowledges support by the ARC Discovery Project DP140100648. 

\section*{References}
\bibliography{reply_Gillis_v21_resub}

\end{document}